\documentclass[11pt,letter]{article}
\usepackage{auxlib}
\pdfoutput=1 

\usepackage{auxlib}
\usepackage{authblk}
\usepackage{latexsym}

\title{Constant Approximation for Weighted Nash Social Welfare with Submodular Valuations}

\author[1]{Yuda Feng \thanks{The work of Yuda Feng and Shi Li was supported by the State Key Laboratory for Novel Software Technology and the New Cornerstone Science Laboratory.}}
\author[2]{Yang Hu }
\author[1]{Shi Li $^\ast$}
\author[3]{Ruilong Zhang }

\affil[1]{\small School of Computer Science, 
Nanjing University, Nanjing, China}
\affil[2]{\small Institute for Interdisciplinary Information Sciences, Tsinghua University, Peking, China}
\affil[3]{\small Department of Mathematics, Technical University of Munich, Munich, Germany}

\affil[ ]{\texttt{yudafeng@smail.nju.edu.cn, y-hu22@mails.tsinghua.edu.cn, shili@nju.edu.cn, ruilong.zhang@tum.de}}

\date{}

\begin{document}

\maketitle

\begin{abstract}
We study the problem of assigning items to agents so as to maximize the \emph{weighted} Nash Social Welfare (NSW) under submodular valuations.
The best-known result for the problem is an $O(nw_{\max})$-approximation due to Garg, Husic, Li, Végh, and Vondrák~[STOC 2023], where $w_{\max}$ is the maximum weight over all agents. 
Obtaining a constant approximation algorithm is an open problem in the field that has recently attracted considerable attention.

We give the first such algorithm for the problem, thus solving the open problem in the affirmative. 
Our algorithm is based on the natural Configuration LP for the problem, which was introduced recently by Feng and Li~[ICALP 2024] for the additive valuation case. 
Our rounding algorithm is similar to that of Li~[SODA 2025] developed for the unrelated machine scheduling problem to minimize weighted completion time. 
Roughly speaking, we designate the largest item in each configuration as a large item and the remaining items as small items. 
So, every agent gets precisely 1 fractional large item in the configuration LP solution. 
With the rounding algorithm in Li~[SODA 2025], we can ensure that in the obtained solution, every agent gets precisely 1 large item, and the assignments of small items are negatively correlated.
    
\end{abstract}

\newpage

\section{Introduction}

We study the problem of allocating a set $M$ of indivisible items among a set $N$ of agents, where each agent $i\in\agents$ has a monotone non-negative submodular valuation $v_i:2^{M}\to \R_{\geq 0}$ and a weight $w_i\in(0,1)$ with $\sum_{i\in\agents}w_i=1$.
The {\em weighted} Nash Social Welfare (NSW) problem under submodular valuations asks for partition $\cS:=(S_i)_{i\in\agents}$ of $\items$ that maximizes the weighted geometric mean of the agents' valuations:
\[
\mathsf{NSW}(\cS) = \prod_{i\in\agents}\left(v_i(S_i)\right)^{w_i}.
\]
The case when all $w_i$'s are equal to $1/n$ is called the unweighted Nash Social Welfare problem. 
As usual, we assume we are given a value oracle for each $v_i$.  
W.l.o.g, we assume $v_i(\emptyset) = 0$ for every agent $i \in \agents$ \footnote{If $v_i(\emptyset) > 0$ for some $i \in \agents$, we can create a ``private'' item $j_i$ for $i$ which has $0$-value to all agents other than $i$. We replace the valuation of $i$ with $v'_i$, which is defined as follows: $v'_i(S) := v_i(S) - v_i(\emptyset)$ if  $j_i \notin S$ and $v'_i(S) = v_i(S \setminus j_i)$ if $j_i \in S$.}.
    
Fair and efficient allocation of resources is a central problem in computer science, game theory, and social choices, with applications across diverse domains~\cite{BT05,BT96,BCE16,KN79,Mou04,RW98,Rot15,You94}.
Three distinct communities independently discovered the notation of Nash social welfare: as a solution to the bargaining problem in classical game theory~\cite{Nas50}, as a well-established concept of proportional fairness in networking \cite{Kel97}, and as the celebrated notion of competitive equilibrium with equal incomes in economics \cite{Var74}.
The unweighted case for the problem was introduced by Nash~\cite{Nas50}, and it was later extended to the weighted case \cite{HS72, Kal77}. 
This extension has since been widely studied and applied across various fields, such as bargaining theory \cite{CM10, LV07, Tho86}, water allocation \cite{DWL+18,HdLGY14}, climate agreements \cite{YvIWZ17}, and more.
One of the most important features of the NSW objective is that it offers a tradeoff between the frequently conflicting demands of fairness and efficiency.

A special case for the valuations $v_i$ is when they are additive. 
The unweighted NSW problem with additive valuations is an important topic in optimization and has received considerable interest. 
Barman, Krishnamurthy, and Vaish \cite{BKV18} developed a \((\ce^{1/\ce}\approx 1.445)\)-approximation algorithm that finds an allocation that is both Pareto-efficient and envy-free up to one item (EF1). 
They showed that this problem can be reduced to the case of identical valuations, where any EF1 allocation can achieve an approximation ratio of \(\ce^{1/\ce} \approx 1.445\).
On the negative side, Garg, Hoefer, and Mehlhorn \cite{GHM23} established a hardness of \(\sqrt{8/7}\).

For the weighted case with additive valuations, Brown, Laddha, Pittu, and Singh~\cite{BLP24} introduced an approximation algorithm with a ratio of \(5\cdot \text{exp}(2\log n + 2\sum_{i\in A} w_i \log w_i)\). 
Later, Feng and Li \cite{FL24} presented an elegant $(\ce^{1/\ce}+\epsilon)$-approximation algorithm for the weighted case, using their novel configuration LP and the Shmoys-Tardos rounding procedure developed in the context of unrelated machine scheduling.  
The approximation ratio matches the best-known ratio for the unweighted case.

When the $n$ valuation functions are additive and identical, Nguyen and Rothe~\cite{NR14} developed a PTAS for the unweighted NSW problem.
Later, Inoue and Kobayashi~\cite{IK22} gave an additive PTAS for the problem, i.e., a polynomial-time algorithm that maximizes the Nash social welfare within an additive error of $\epsilon v_{\max}$,
where $v_{\max}$ is the maximum utility of an item.
    
Li and Vondrák~\cite{LV22} developed the first constant approximation algorithm for unweighted NSW with submodular valuations using convex programming. 
The ratio has been improved by Garg, Husic, Li, Végh, and Vondrák \cite{GHL23} to $(4+\epsilon)$ using an elegant local-search-based algorithm. 
When additionally $n = O(1)$, by guessing the value and the $O(1)$ largest items for each agent, and using the multilinear extension of submodular functions, a $\ce/(\ce-1)$-approximation can be achieved \cite{GKK23}.
In the same paper, \cite{GKK23} proved that unweighted NSW with submodular valuations is hard to approximate within $\ce/(\ce-1)-\epsilon$. 
The hardness holds even for the case $n = O(1)$.

For the weighted NSW problem with submodular valuations, \cite{GHL23} showed that the approximation ratio of the local search algorithm becomes \(O(nw_{\max})\), where $w_{\max} := \max_{i \in [n]} w_i$ is the maximum weight over the agents.  
In the new version \cite{GHL24} of the paper, the authors presented a $(6\ce+\epsilon)$-approximation algorithm with running time $2^{O(n\log n)}\poly(m,1/\epsilon)$, which is polynomial when $n = O(1)$.

For the more general setting where the valuations are subadditive, Dobzinski, Li, Rubinstein, and Vondrák \cite{DLR23} recently proposed a constant approximation algorithm when agents are unweighted, provided that we have access to demand oracles for the valuation functions. 

\paragraph{Our Result.} 
In this paper, we give the first polynomial-time $O(1)$-approximation algorithm for weighted Nash social welfare under the submodular valuations.  
The best result prior to this work was the $O(nw_{\max})$-approximation due to Garg, Husic, Li, Végh, and Vondrák \cite{GHL23}.
\begin{theorem}
For any $\epsilon>0$, there is a randomized $(233+\epsilon)$-approximation algorithm for the weighted Nash social welfare problem with submodular valuations, with running time polynomial in the size of the input and $\frac{1}{\epsilon}$.
\label{thm:ratio}
\end{theorem}

For convenience, we list the known approximation results for the NSW problem in \cref{tab:summary}.

\begin{table}[htb]\centering
        \begin{tabular}{c|c|c|c|c|c|c}
            \hline  & \multicolumn{2}{|c|}{Additive} & \multicolumn{2}{|c|}{Submodular} & \multicolumn{2}{|c}{Subadditive} \\\hline
             & LB & UB & LB & UB & LB & UB \\\hline
            Unweighted & $\sqrt{\frac{8}{7}}$~\cite{GHM23} & $\ce^{1/\ce}+\epsilon$~\cite{BKV18} & $\frac{\ce}{\ce-1}$~\cite{GKK23} & $4+\epsilon$~\cite{GHL23} & & $O(1)^*$~\cite{DLR23} \\\hline
            Weighted  & &  $\ce^{1/\ce}+\epsilon$~\cite{FL24} & & $233+\epsilon$ (\cref{thm:ratio}) & & \\\hline
        \end{tabular}
        \caption{Known Results for Nash social welfare. LB and UB stand for lower and upper bounds, respectively. The result with $^*$ requires demand oracles for valuation functions. When the function is identical additive, the upper bound for the unweighted case is PTAS~\cite{IK22,NR14}. When $n = O(1)$, the upper bounds for unweighted and weighted NSW with submodular valuations are respectively $\ce/(\ce-1)$~\cite{GKK23} and $6\ce+\epsilon$~\cite{GHL24}.
        }
        \label{tab:summary}
    \end{table}

\subsection{Overview of Our Techniques}
Our algorithm leverages the configuration LP introduced in \cite{FL24} for the additive valuation case. 
For each agent \( i \in \agents \) and subset of items \( S \subseteq \items \), we define a variable \( y_{i,S} \) to indicate whether the set of items assigned to \( i \) is precisely \( S \). 
The objective of this LP is to minimize \( \sum_{i, S} y_{i, S} \cdot w_i \cdot \ln v_i(S) \), the logarithm of the NSW objective.
After solving the LP, we apply the rounding procedure from \cite{Li25}, developed for the weighted completion time minimization problem in the unrelated machine scheduling setting. 
Then we prove concentration bounds for the values obtained by each agent, using arguments developed for pipage rounding. 

To build intuition, let us focus on the unweighted case. 
For the special case where \( |M| = |N| \), the problem reduces to a maximum-weight bipartite matching problem with weights given by the logarithm of values. 
So, any general algorithm for the problem must capture the maximum weight of the bipartite matching algorithm as a special case.  
Interestingly, previous results showed that if one is given the largest (i.e., the most valuable) item assigned to every agent, then an $O(1)$-approximation algorithm is easy to obtain using local search \cite{GHL23} or LP rounding \cite{GHV21,LV22}. 
For example, with this idea, Garg, Husic, Li, Végh, and Vondrák \cite{GHL23} designed an elegant 4-approximation local search algorithm. They first compute an initial matching of one item to every agent so as to maximize the NSW objective, then assign the remaining items using local search with an endowed valuation function, and finally rematch the initially assigned items to agents to maximize the final Nash social welfare. Unfortunately, their algorithm fails to give an \( O(1) \)-approximation when the agents are weighted. 

Our algorithm implements the idea of ``matching largest items to agents'' using the configuration LP solution as a guide. We achieve a per-client guarantee, allowing us to give an $O(1)$-approximation for the \emph{weighted} NSW problem with submodular valuations. After obtaining an LP solution \( (y^*_{i, S})_{i, S} \), for each agent \( i \) and configuration \( S \), we designate the largest item in \( S \) as a ``large'' item for $i$, while treating the remaining items as ``small''. This creates a fractional assignment in which each agent receives exactly one fractional large item. While maintaining marginal probabilities in our rounding algorithm, we ensure that each agent gets exactly one large item, and the assignment of small items are negatively correlated. That is, we select a random matching for large items. 

If the large and small items were disjoint, the rounding algorithm would be straightforward. 
However, complications arise when an item may be large for one agent and small for another — or even for the same agent in different configurations. This necessitates a correlated assignment strategy for large and small items. This is where we employ the iterative rounding procedure of \cite{Li25}. We construct a bipartite multi-graph between agents and items, with two edge types: \emph{marked} edges for large items and \emph{unmarked} edges for small items. During iterative rounding, we identify either a simple cycle of marked edges or a pseudo-marked path -- a simple path of marked edges with two unmarked edges at the ends -- and apply rotation or shifting operations on the cycle or path in each iteration. This process ultimately yields an integral assignment.

To analyze the approximation ratio, we focus on each agent \( i \) and analyze \( \E[\ln(v_i(T))] \), where \( T \) is the set of items assigned to \( i \). Note that \( T \) includes exactly one large item, respecting the marginal probabilities. Let \( T^\mathrm{S} \) denote the remaining items, i.e., the small items assigned to \( i \). The assignments of the large item and the small items may be positively correlated, so we analyze the worst-case scenario for this correlation. However, the assignments of the small items are negatively correlated; more precisely, they are determined through a pipage-rounding procedure. Using the concave pessimistic estimator technique from \cite{HO14} and the submodularity of the function \( v_i \), we can establish concentration bounds for \( v_i(T^\mathrm{S}) \). With the bounds, we can lower bound \( \E[\ln(v_i(T))] \) by $\sum_{S}y^*_{i,S} \ln v_i(S) - O(1)$.

\paragraph{Organization.} The rest of the paper is organized as follows. We introduce some preliminaries in \cref{sec:prelim}, describe our algorithm in \cref{sec:alg}, and give its analysis in \cref{sec:analysis}. For a smoother flow in the main text, we defer some proofs to the appendix.
\section{Preliminaries}
\label{sec:prelim}

\begin{definition}[Monotone Submodular Functions]
Let $\items$ be the ground set (item set) and $f:2^{M}\to\R_{\geq 0}$ be a function defined over $\items$.
\begin{itemize}
    \item $f$ is submodular if for any $S,T\subseteq M$, $f(S)+f(T) \geq f(S\cap T) + f(S\cup T)$.
    \item $f$ is monotone if $f(S)\leq f(T)$ whenever $S\subseteq T$.
\end{itemize}
\end{definition}
For convenience, we slightly abuse the notation and use $f(j)$ to represent $f(\set{j})$.

\subsection{Truncations of Submodular Functions}
\begin{definition}[Truncated Function]
    \label{def:truncate}
    Given a monotone submodular function $f: 2^M \to \R_{\geq 0}$ with $f(\emptyset) = 0$ and a real $R > 0$, define $f^{(R)}: 2^M \to \R_{\geq 0}$ to be the following function:
    \begin{align*}
        f^{(R)}(S): = \E_{\theta \sim [0, 1]}\left[ f\left(\Big\{j \in S: \theta \leq \frac{R}{f(j)}\Big\}\right) \right], \quad \forall S \subseteq M.
    \end{align*}
    We say $f^{(R)}$ is the function obtained from $f$ by truncating individual values by $R$. 
\end{definition}

The truncated function provides a trackable lower bound for the valuation function. After truncating the function, the marginal values will be bounded by $R$; this allows us to apply the concentration bound. 
This is given in the following lemma:
\begin{lemma}\label{lem:truncate}
Let $f: 2^M \to \R_{\geq 0}$ be a monotone submodular function and a $R > 0$ be a real. Let $R' \geq R$ be a real and $S \subseteq M$ be a non-empty set such that $\max_{j \in S}f(j) \leq R'$.
Then, the following properties are true:
\begin{enumerate}[label=(\ref{lem:truncate}\ablue{\alph*}),leftmargin=*,align=left]
    \item $f^{(R)}$ is a monotone submodular function such that $f^{(R)}(j) \leq R$ for all $j\in M$.
    \label{prop:trun:submodular}
    \item $f(S) \geq f^{(R)} (S) \geq \frac{R}{R'}\cdot f(S)$.
    \label{prop:trun:lower-bound}
\end{enumerate}   
\end{lemma}

\begin{proof}
It is easy to see that $f^{(R)}$ is a monotone function.
The truncated function $f^{(R)}$ can be regarded as a linear combination of submodular functions, and thus, it is also a submodular function.
For any $j \in M$, we have $f^{(R)}(j) = \min\{\frac{R}{f(j)}, 1\} \cdot f(j) \leq R$. Thus, \ref{prop:trun:submodular} holds.

For \ref{prop:trun:lower-bound}, the proof of $f(S)\geq f^{(R)}(S)$ is straightforward.
If $\theta\leq\frac{R}{R'}$, then $\set{j\in S: \theta \leq \frac{R}{f(j)}}=S$ as $f(j)\leq R'$ for all $j\in S$.
Thus, $f^{(R)}(S)\geq \frac{R}{R'}\cdot f(S)$.
This finishes the proof of \cref{lem:truncate}.
\end{proof}

We remark that when the valuation function is additive, the definition of the truncated function can be simplified: $f^{(R)}(S):=\sum_{j\in S}\min\{ R,f(j) \}$. \medskip

\subsection{Extensions of Submodular Functions}
We shall also consider the multilinear and concave (maximum) extensions of a submodular function, which are widely used in the literature.
The concave extension is also known as the ``Configuration LP'' extension in the literature.
\begin{definition}[Multilinear Extension]
Given a monotone submodular function $f:2^M\to \R_{\geq 0}$, its multilinear extension $F:[0,1]^M \to \R_{\geq 0}$ is defined as:
\[
F(\bx) = \sum_{S\subseteq M} f(S) \prod_{j\in S} x_j \prod_{j\in (M\setminus S)}(1-x_j).
\]
\end{definition}

\begin{definition}[Concave Extension]
Given a monotone submodular function $f:2^M\to\R_{\geq 0}$, its concave extension $f^+:[0,1]^M\to\R_{\geq 0}$ is:
\[
f^+(\bx) = \max\left\{ \sum_{S\subseteq M} \alpha_S f(S): \sum_{S\subseteq M} \alpha_S \leq 1;\forall S\subseteq M, \alpha_S\geq 0; \forall j\in M, \sum_{S:j\in S} \alpha_S = x_j  \right\}.
\]
\label{def:concave}
\end{definition}

The following relation between the multilinear and concave extensions has been shown in \cite[Lemma 3.7, 3.8]{vondrak2007submodularity}.
\begin{lemma}[\cite{vondrak2007submodularity}]
Given a monotone non-negative submodular function $f$, its multilinear extension $F$, and its concave extension $f^+$, and a point $\bx\in[0,1]^M$, we have 
$f^{+}(\bx) \geq F(\bx) \geq (1-\frac{1}{\ce})\cdot f^+(\bx)$.
\label{lem:concave-multilinear}
\end{lemma}

\subsection{Concentration Bound for Submodular Functions in Pipage Rounding}

We consider the following modified version of the pipage rounding procedure described in \cref{alg:pipage}. 
This enables us to use the Chernoff-type concentration bounds stated in \cref{thm:pipage-rounding}. 
Suppose we are guaranteed that the algorithm terminates in the finite number of iterations with probability 1.  

\begin{algorithm}[H]
    \caption{Modified Pipage Rounding}
    \begin{algorithmic}[1]
        \Require{$\bx^* \in [0, 1]^{\bar n}$.}
        \Ensure{an integral $\bx \in \{0, 1\}^{\bar n}$.}
        \State $\bx \gets \bx^*$.
        \While{$\bx$ is not integral}
            \State do one of the two operations arbitrarily:
            \Statex\hspace*{\algorithmicindent}{\bf Operation 1:}
            \State\hspace*{\algorithmicindent}choose one coordinate $a \in [\bar n]$ with $x_a \in (0, 1)$, two reals $\delta_1 \in (0, x_a]$ and $\delta_2 \in (0, 1- x_a]$.
            \State\hspace*{\algorithmicindent}with probability $\frac{\delta_2}{\delta_1 + \delta_2}$ \textbf{do}: $x_a \gets x_a - \delta_1$,  \textbf{else do}: $x_a \gets x_a + \delta_2$.
            \Statex\hspace*{\algorithmicindent}{\bf Operation 2:}
            \State\hspace*{\algorithmicindent}choose two distinct coordinates $a, b \in [\bar n]$ with $x_a, x_b \in (0, 1)$, two reals $\delta_1 \in (0, \min\{x_a, 1 - x_b\}]$, and $\delta_2 \in (0, \min\{1 - x_a, x_b\}]$.
            \State\hspace*{\algorithmicindent}with probability $\frac{\delta_2}{\delta_1 + \delta_2}$ \textbf{do}: $x_a \gets x_a - \delta_1, x_b \gets x_b + \delta_1$.
            \State\hspace*{\algorithmicindent}\hspace*{75pt} \textbf{else do}: $x_a \gets x_a + \delta_2, x_b \gets x_b - \delta_2$.
      \EndWhile
      \State \Return $\bx$
    \end{algorithmic}
        \label{alg:pipage}
\end{algorithm}

We shall use $\bx$ to denote the final $\bx$ vector returned by \cref{alg:pipage}.  
It can be shown that $x_j$'s are negatively correlated following a similar argument to that used in the pipage rounding procedure by \cite{CVZ10}. 
However, it is unknown whether the negative correlation alone suffices to establish the Chernoff-type concentration bounds we need. 
Therefore, we provide a direct proof of the concentration bound for $\bx$ using the concave estimator technique from~\cite{HO14}.
We prove the following theorem in \cref{sec:pipage-rounding}:
\begin{theorem}
    \label{thm:pipage-rounding}
    Let $v:2^{[\bar n]} \to \R_{\geq 0}$ be a monotone submodular function with marginal values at most 1, and $F:[0, 1]^{\bar n} \to \R_{\geq 0}$ be the multilinear extension of $v$. Let $\bx^* \in [0, 1]^{\bar n}$, $\mu = F(\bx^*)$, $\bx \in \{0, 1\}^{\bar n}$ be the output of \cref{alg:pipage} for the input $\bx^*$, and $U = \{i \in [\bar n]: x_i = 1\}$.
    Then, for any $\delta \in (0, 1)$, we have 
    \begin{align*}
        \Pr[v(U) \leq (1-\delta)\mu] \leq \ce^{-\delta^2\mu/2}.
    \end{align*}
\end{theorem}

\section{Iterative Rounding for Weighted Submodular Nash Social Welfare using Configuration LP}
\label{sec:alg}

In this section, we give our algorithm for the weighted submodular NSW problem. 
We describe the configuration LP relaxation in \cref{subsec:LP}, the construction of the bipartite multi-graph $G$ in \cref{subsec:G}, and the iterative rounding algorithm in \cref{subsec:rounding} respectively.

\subsection{LP Formulation}
\label{subsec:LP}

We start with the configuration LP relaxation \eqref{Conf-LP} for our problem, which is the same as the one used in \cite{FL24}, except now, each $v_i$ is a submodular function.  
In the correspondent integer program, we have a variable $y_{i,S}\in\set{0,1}$ for every agent $i\in\agents$ and set $S\subseteq \items$ indicating if the set of items assigned to agent $i$ is precisely $S$ or not. 
The objective is to maximize the logarithm of the weighted Nash social welfare, which is $\sum_{i\in \agents,S\subseteq \items} w_i \cdot  y_{i,S} \cdot \ln(v_i(S))$. 
\eqref{LPC:item} ensures that each item is assigned to exactly one agent.
\eqref{LPC:agent} ensures that each agent is given precisely one item set.

\begin{align}
    \text{max} && \sum_{i\in \agents,S\subseteq \items} w_i \cdot  y_{i,S} \cdot \ln(v_i(S)) & \tag{\text{Conf-LP}} \label{Conf-LP}\\
    \text{s.t.} \nonumber \\
    &&\sum_{S:j\in S}\sum_{i\in \agents}y_{i,S} &= 1, &\forall j\in \items \label{LPC:item} \\
    &&\sum_{S\subseteq \items}y_{i,S} &= 1, &\forall i\in \agents \label{LPC:agent}\\
    &&y_{i,S} &\geq 0, &\forall i\in \agents, S\subseteq \items 
\end{align}

There are an exponential number of variables in the LP. 
Using standard techniques, we can consider the dual of the LP and design a separation oracle for it. 
However, unlike the additive valuation case, for which the separation oracle incurs an additive $\epsilon$ error, the error becomes $\ln (\frac{\ce}{\ce - 1} + \epsilon)$ for the submodular valuation case (this corresponds to a multiplicative factor of $(\frac{\ce}{\ce-1} + \epsilon)$ for the weighted NSW objective).
This comes from the approximation factor for the problem of maximizing a monotone submodular function with a knapsack constraint~\cite{SV04}.  
The proof of the lemma is deferred to \cref{sec:oracle}: 
\begin{lemma}
For any constant $\epsilon > 0$, the Configuration LP \eqref{Conf-LP} can be solved in polynomial time within an additive error of $\ln (\frac{\ce}{\ce-1}+\epsilon)$. 
\label{lem:oracle}
\end{lemma}

\subsection{Construction of Bipartite Multi-Graph $G = (\agents \cup \items, E)$ with Marked and Unmarked Edges} \label{subsec:G}
After we solve \eqref{Conf-LP}, we obtain a solution $\by^*:=(y^*_{i,S})_{i\in \agents,S\subseteq \items}$, represented using the list of non-zero coordinates. 
From now on, we say $S \subseteq \items$ is a configuration for an agent $i$ if $y^*_{i, S} > 0$.

\begin{definition}[Large and Small Items]
For each agent $i$ and configuration $S$ for $i$, let $\kappa^{(i)}_S$ be the largest item in $S$, i.e., $\kappa^{(i)}_S:=\argmax_{j\in S}v_i(j)$. 
We say $j$ is a large item for $i$ if $j = \kappa^{(i)}_S$ for some configuration $S$ for $i$. 
Let $\largeitems[i]$ be the set of large items for agent $i$.
We say $j$ is a small item for $i$ if $j \in S \setminus \kappa^{(i)}_S$ for some configuration $S$ for $i$. 
Let $\smallitems[i]$ be the set of small items for $i$. 
\end{definition}

Note that an item may be both a large and small item for an agent $i$. 
It may also happen that for a small item $j$, and a large item $j'$ for an agent $i$, we have $v_i(j) > v_i(j')$.  \medskip

Our rounding algorithm is similar to that of \cite{Li25} for the unrelated machine weighted completion time problem. 
We build a bipartite multi-graph $G:=(\agents \cup \items, E)$ between agents $\agents$ and items $\items$, where each edge in $E$ is either {\em marked} or {\em unmarked}. 
We also define a vector $\bx^*$ over the edges. 
For each agent $i$ and item $j \in \largeitems[i]$, we create a marked edge $ij$ with $x^*$-value $x^*_{ij}:=\sum_{S:\kappa^{(i)}_S=j}y^*_{i,S}$. 
For each agent $i$ and item $j \in \smallitems[i]$, we create an unmarked edge with $x^*$-value $x^*_{ij}:=\sum_{S:j\in S\setminus \kappa^{(i)}_S} y^*_{i,S}$. 
So, there might be two parallel edges $ij$ between an agent $i$ and an item $j$; in this case, one of them is marked, and the other is unmarked. 
From now on, when we refer to an edge $ij$, we assume we know its identity, which will decide if $j$ is large or small. 

\begin{figure}[htb]
    \centering
    \includegraphics[width=14cm]{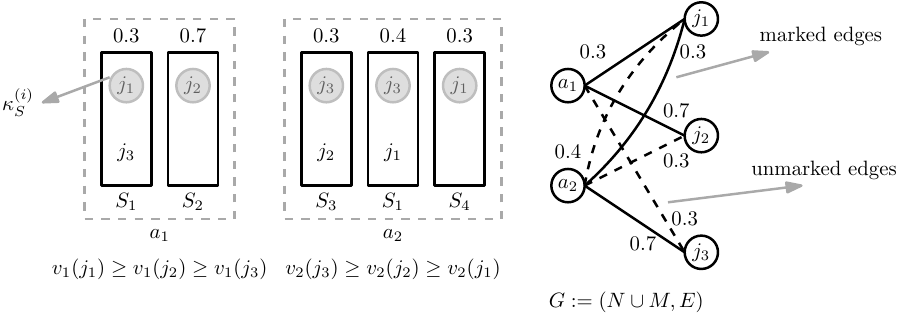}
    \caption{Illustration for the constructed bipartite multi-graph. There are two agents and three items, and the agents' preference for items is shown in the figure. Suppose that, after solving \eqref{Conf-LP}, we obtain $y^*_{1,S_1}=0.3,y^*_{1,S_2}=0.7$, $y^*_{2,S_3}=0.3,y^*_{2,S_1}=0.4,y^*_{2,S_4}=0.3$. By agents' preference, we know that $\largeitems[1]=\set{j_1,j_2}$ and $\largeitems[2]=\set{j_1,j_3}$, they are marked by gray cycle. The small item set is $\smallitems[1]=\set{j_3},\smallitems[2]=\set{j_1,j_2}$. Then, we can create a bipartite multi-graph as stated above. In the right part of the figure, we use a solid line to represent the marked edges (edges between agents and large items) and a dashed line to represent the unmarked edges (edges between agents and small items). The example shows two edges between agent $a_2$ and item $j_1$; thus, the bipartite graph can be a multi-graph. Verifying the two properties of the constructed vector $\bx^*$ is also easy. }
    \label{fig:bipartite}
\end{figure}

Notice that the constructed vector $\bx^*$ has the following properties: 
\begin{enumerate}[label=(\roman*)]
    \item For any item $j \in \items$, we have $\sum_{ij \in E}x^*_{ij}=1$. 
    \item For any agent $i \in \agents$, we have $\sum_{j\in \largeitems[i]}x^*_{ij}=1$.
\end{enumerate}
The first property is due to Constraint \eqref{LPC:agent} of \eqref{Conf-LP}, and the second is due to Constraint \eqref{LPC:item} of \eqref{Conf-LP}. 
An example can be found in \cref{fig:bipartite}.

\subsection{Rounding $\bx$ into an Integral Solution} \label{subsec:rounding}

Our iterative rounding algorithm is also similar to that of \cite{Li25}. 
During the algorithm, we maintain a vector $\bx \in [0, 1]^E$, which is set to $\bx^*$ initially. 
In each round, we find either a \emph{marked cycle} or a \emph{pseudo-marked path} in $\fsupp(\bx)$, where we shall use $\fsupp(\bx) := \set{ij \in E: x_{ij} \in (0, 1)}$ to denote the set of edges in $E$ with strictly fractional $x$ values.  
We apply the rotation/shifting operation to the structure we found until $\bx$ becomes integral. 
The operations we use here are slightly simpler than those in \cite{Li25} as we maintain the total $x$-values of marked edges incident to an agent, while the algorithm in \cite{Li25} needs to maintain the total ``volume'', which makes the procedure more complicated. 

First, we define \emph{marked cycles} and \emph{pseudo-marked paths}.
In the remainder of this subsection, we use $i$ and $j$ to represent an agent and an item, respectively.

\begin{definition}[Marked Cycle and Pseudo-Marked Path] \label{def:cycle-path}
Given a spanning sub-graph $G' = (\agents \cup \items, E')$ of $G$, a marked cycle in $G'$ is a simple cycle in $G'$ that consists of marked edges only. A pseudo-marked path $(i_1,j_1,\ldots,j_k,i_{k+1}), k \geq 1$ is a (not-necessarily-simple) path of $G'$ such that 
\begin{enumerate}[label=(\ref{def:cycle-path}\ablue{\alph*}),leftmargin=*]
    \item the subpath $(j_1,\ldots,j_k)$ is a simple path that consists of marked edges only, and 
    \item the edges $(i_1,j_1)$ and $(j_k,i_{k+1})$ are distinct unmarked edges.
\end{enumerate}
\end{definition}

For a pseudo-marked path, we do not impose any other conditions that are not stated in the definition. 
For example, $i_1$ and/or $i_{k+1}$ may be the same as some agent in $\set{i_1, i_2, \ldots, i_{k-1}}$; in this case, the pseudo path shall contain cycles with both marked and unmarked edges. 
It may also happen that $i_1 = i_{k+1}$. 
It is also possible that $k = 1$, in which case the pseudo path consists of two distinct unmarked edges, which implies $i_1 \neq i_2$.

Given any vector $\bx\in[0,1]^{E}$, we define $G(\bx) := (V, \fsupp(\bx))$ to be the spanning graph of $\bx$ containing the edges with $x$-values strictly between $0$ and $1$.  
The iterative rounding algorithm is described in \cref{alg:rounding}. 

\begin{algorithm}[htb] 
\caption{Iterative Rounding Algorithm for $\bx^*$}
\label{alg:rounding}
\begin{algorithmic}[1]
\Require the bipartite multi-graph $G:=(\agents\cup \items, E)$ and the fractional vector $\bx^*:=(x^*_e)_{e\in E}$.
\Ensure an integral solution $\bx$.
\State $\bx \gets \bx^*$.
\While{there exists a variable $x_e$ such that $0<x_e<1$}
\State find either a marked cycle $(i_1, j_1, i_2, j_2, \ldots, i_k, j_k, i_{k+1} = i_1)$ or a pseudo-marked path $(i_1, j_1, i_2, j_2, \ldots, i_k, j_k, i_{k+1})$ in $G(\bx)$. \label{State:cycle-path}
\smallskip
\State $\displaystyle \delta_1\gets\min\Big\{\min_{a \in [k]} x_{i_aj_a},\ \min_{a \in [k]}(1-x_{j_ai_{a+1})}\Big\}, \qquad \displaystyle \delta_2\gets\min\Big\{\min_{a \in [k]}(1 - x_{i_aj_a}),\ \min_{a \in [k]}x_{j_ai_{a+1}}\Big\}$.
\State with probability $\displaystyle \frac{\delta_2}{\delta_1+\delta_2}$ \textbf{do}: \textbf{for} every $a \in [k]$ \textbf{do}: $x_{i_aj_a} \gets x_{i_aj_a} - \delta_1, x_{j_ai_{a+1}} \gets x_{j_ai_{a+1}} + \delta_1$. \smallskip
\State \hspace{3.08cm} else \textbf{do}: \textbf{for} every $a \in [k]$ \textbf{do}: $x_{i_aj_a} \gets x_{i_aj_a} + \delta_2, x_{j_ai_{a+1}} \gets x_{j_ai_{a+1}} - \delta_2$.
\EndWhile
\State \Return $\bx$.
\end{algorithmic}
\end{algorithm} 

We first assume that in Step~\ref{State:cycle-path}, we can always find a marked cycle or a pseudo-marked path. Then, it is easy to see that we maintained the following two invariants 
\begin{enumerate}[label = (\ablue{I\arabic*})]
    \item $\sum_{ij \in E}x_{ij} = 1$ for every item $j \in \items$.
    \label{invar:item}
    \item $\sum_{j \in \largeitems[i]} x_{ij} = 1$ for every agent $i \in \agents$.
    \label{invar:large-item}
\end{enumerate}
Then, the algorithm terminates in polynomial time as the $x$-value of at least one edge becomes integral in each iteration. 

It remains to show that the goal in Step~\ref{State:cycle-path} can be achieved: 
\begin{lemma}\label{lem:correctness}
    Let $\bx$ be the vector before Step~\ref{State:cycle-path} in some iteration of the while loop; so $\bx$ is not integral.  Then, $G(\bx)$ contains a marked cycle or a pseudo-marked path. 
    \end{lemma}
\begin{proof}
If the marked edges in $G(\bx)$ contain a cycle, then we are done. Assume this does not happen; so the marked edges form a forest. 

If there is at least one marked edge in $G(\bx)$, then we can take a non-empty path of marked edges between two leaf vertices in the forest.  
Due to Invariant (I2), both leaves must be items. 
So, let $j_1, i_2, j_2, i_3, \ldots, i_k, j_k$ be the non-empty simple path of marked edges; $j_1$ and $j_k$ are leaves in the forest.  
Due to Invariant (I1), there must be an unmarked edge incident to $j_1$ in $G(\bx)$; the same holds for $j_k$.  
So, concatenating the two unmarked edges and the path $j_1, i_2, j_2, i_3, \ldots, i_k, j_k$ gives us a pseudo-marked path. 

Now suppose there are no marked edges in $G(\bx)$. As $\bx$ is not integral, there is some item $j$ which is not integrally assigned in $\bx$. 
Then $j$ must be incident to two distinct unmarked edges due to Invariant (I1). The two unmarked edges form a pseudo-marked path. 
\end{proof}

For every agent $i$, let $\bx^{i, \umk}$ be the vector $\bx$ restricted to the set of unmarked edges incident to $i$. 
If we only focus on how $\bx^{i, \umk}$ changes, then \cref{alg:rounding} falls into the algorithmic template stated in \cref{alg:pipage}. 
In particular, if in an iteration we rotate a marked cycle, then $\bx^{i, \umk}$ is unchanged.  
Suppose we shift a pseudo-marked-path $(i_1, j_1, \ldots, i_k, j_k, i_{k+1})$ in an iteration. 
If $i \notin \{i_1, i_{k+1}\}$, then $\bx^{i, \umk}$ is unchanged; if $i = i_1, i \neq i_{k+1}$ or $i \neq i_1, i = i_{k+1}$, then Operation 1 is performed in the iteration in \cref{alg:rounding}; if $i = i_1 = i_{k+1}$, then Operation 2 is performed.  Thus, we can apply the concentration bound over $\bx^{i, \umk}$ using \cref{thm:pipage-rounding}. 

\section{Analysis of the Approximation Ratio} \label{sec:analysis}
In this section, we analyze the approximation ratio achieved by the algorithm.  
For every $e \in E$, let $X_e \in \{0, 1\}$ be the value of $x_e$ returned by \cref{alg:rounding}.  
We first summarize some of the properties we have for the random variables $X_e$'s.

\begin{observation}
\label{obs:alg-probability}
The following properties hold:
\begin{enumerate}[label=(\ref{obs:alg-probability}\ablue{\alph*}),leftmargin=*,align=left]
    \item For each $e\in E$, we have $\E[X_e] = x^*_e$.
    \label{prop:marginal}
    \item For each item $j$, we have $\sum_{ij\in E} X_{ij}=1$ with probability 1.
    \label{prop:item-assigned}
    \item For each agent $i$, we have $\sum_{j\in \largeitems[i]} X_{ij}=1$ with probability 1. 
    \label{prop:large-edge}
\end{enumerate}   
\end{observation} 
\begin{proof}
Fix an edge $e \in E$. Focus on an iteration of the loop in \cref{alg:rounding}. 
Let $x^{\mathrm{old}}_e$ and $x^{\mathrm{new}}_e$ be the value of $x_e$ at the beginning and end of the iteration respectively.  
Then it is easy to see that $\E[x^{\mathrm{new}}_e|x^{\mathrm{old}}_e] = x^{\mathrm{old}}_e$. 
Property~\ref{prop:marginal} then follows. 
Property~\ref{prop:item-assigned} and \ref{prop:large-edge} follow from Invariant~\ref{invar:item} and \ref{invar:large-item}, respectively. 
\end{proof}

Till the end of the section, we fix an agent $i \in \agents$ and let $T$ be the set of items assigned to $i$ by our algorithm. 
As $i$ is fixed, we omit subscripts $i$ from most of the notations: we use $v(S), \kappa(S), y^*_S, \largeitems$ and $\smallitems$ to denote $v_i(S), \kappa_i(S), y^*_{i,S}, \largeitems[i]$ and $\smallitems[i]$ respectively.  
Notice that by \ref{prop:large-edge} of \cref{obs:alg-probability}, $T$ contains exactly one large item, i.e., an item in $\largeitems$. 
We use $k^\rmL$ to denote the unique large item in $T$. 
Let $T^\rmS := T \cap \smallitems$ be the set of small items assigned to the agent $i$. 
So, $T = \{k^\rmL\} \uplus T^\rmS$. 
The goal of the section is to prove \cref{lem:analysis-per-agent}.
\begin{lemma}
    \label{lem:analysis-per-agent}
    $\displaystyle \E\big[\ln(v(T))\big] \geq \sum_{S} y^*_{S} \ln v(S) - \left(3+\frac{26}{\ce^3} + \ln 2\right)$.
\end{lemma}

We show how the lemma implies \cref{thm:ratio}.

\begin{proof}[Proof of \cref{thm:ratio}] In this proof, we consider all agents $i \in \agents$ and thus we include the subscript $i$ in all notations. Also, let $T_i$ denote the set of items assigned to the agent $i$ by our algorithm. 
We apply \cref{lem:analysis-per-agent} to each agent $i\in N$, and by the linearity of expectation and $\sum_{i\in N}w_i=1$, we have:
\[
\E\left[ \sum_{i\in N} w_i \cdot \ln v_i(T_i) \right] \geq \sum_{i\in N, S\subseteq M} w_i \cdot y^*_{i,S} \ln v_i(S) - \left( 3+\frac{26}{\ce^3}+\ln 2\right).
\]
Raising both sides to the exponent and applying Jensen's inequality, we have:
\[
\E\left[ \prod_{i\in N}v_i(T_i)^{w_i} \right]\geq \exp\left( - \left( 3+\frac{26}{\ce^3}+\ln 2\right) \right) \cdot \exp\left( \sum_{i\in N, S\subseteq M} w_i \cdot y^*_{i,S} \ln v_i(S) \right).
\]
By \cref{lem:oracle}, we have:
\[
\exp\left( \sum_{i\in N, S\subseteq M} w_i \cdot y^*_{i,S} \ln v_i(S) \right) \geq \opt \cdot \left(1-\frac{1}{\ce}-\epsilon \right).
\]
Thus, we get:
\[
\E\left[ \prod_{i\in N}v_i(T_i)^{w_i} \right]\geq \exp\left( - \left( 3+\frac{26}{\ce^3}+\ln 2\right) \right) \cdot \left(1-\frac{1}{\ce}-\epsilon \right) \cdot \opt.
\]
Therefore, the overall approximation factor is $(\frac{\ce}{\ce-1}+\epsilon)\cdot 2\cdot \exp(3+26/\ce^3) < 233+\epsilon$.
\end{proof}

Now, it remains to show \cref{lem:analysis-per-agent}.
To this end, we first find a lower and upper bound for the value of $\E[\ln v(T)]$ and $\sum_{S}y^*_S \ln v(S)$; see \cref{sec:input-output}.
Then, we shall compare these two bounds in \cref{sec:comparing} to obtain \cref{lem:analysis-per-agent}.

\subsection{Input and Output Distributions}
\label{sec:input-output}

We have two distributions over sets of items: the \emph{input distribution}, which is the distribution of configurations for $i$, and the \emph{output distribution}, which corresponds to the distribution for the set of items assigned to agent $i$ by \cref{alg:rounding}.  
We set up notations for both distributions. 

\paragraph{Notations and Histogram for Input Distribution.} 

In the input distribution, the probability for a set $S$ is $y^*_{S}$; recall that a configuration $S$ for $i$ is a set $S$ with $y^*_{S} > 0$. 
So the total probabilities over all configurations $S$ is $1$ by the LP constraint \eqref{LPC:agent}. 
We define a function $\pi:[0, 1] \to 2^M$ as follows. 
We sort all the configurations $S$ in descending order of $\kappa(S)$, breaking ties arbitrarily.
Let $\prec$ denote this order: $S' \prec S$ if $S'$ appears before $S$ in the order, and $S' \preceq S$ if and only if $S' \prec S$ or $S'=S$. 
Then, $\pi$ is defined as follows: $\pi(t)$ for every $t\in(0,1]$ is the first set $S$ in the order such that $\sum_{S'\preceq S}y^*_{S'} \geq t$. 
Let $\pi(0)$ be the first $S$ in the order with $y^*_S>0$.  
For notational convenience, we use $\pi_t$ to represent $\pi(t)$, and define $\kappa_t := \kappa(\pi_t)$ for every $t \in [0 ,1]$. 
Let $u_t = v(\kappa_t)$ and $B_t = v(\pi_t \setminus \kappa_t)$.

\begin{figure}[htb]
    \centering
    \includegraphics[width=15.5cm]{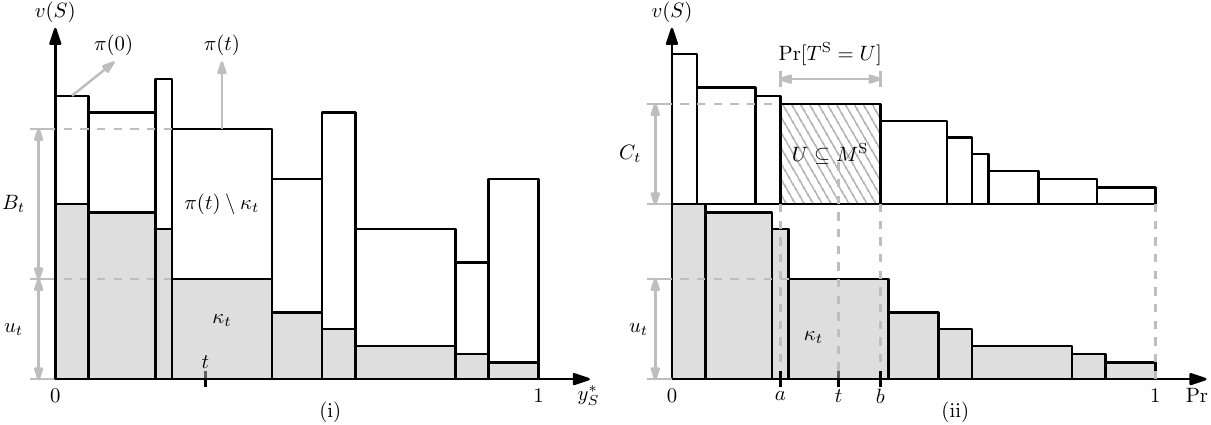}
    \caption{Illustration for the $\pi(\cdot)$ function and $C_t$, which are shown in the subfigure (\rom{1}) and (\rom{2}), respectively.
    For simplicity, the figure considers the case when the valuation function $v$ is additive. 
    In the subfigure (\rom{1}), for each set $S\subseteq M$ with $y^*_S>0$, we have a rectangle. 
    We sort these rectangles in non-increasing order according to the value of the largest item in each set.
    In each rectangle, the gray part is the size of the largest item, and the remaining part is the small item part. 
    In the subfigure (\rom{2}), for each possible algorithm's output of small items, we have a rectangle. 
    We also sort these rectangles in non-increasing order according to the value of each set.
    }
    \label{fig:input_output}
\end{figure}

It is convenient to visualize the notations using a histogram, denoted in the subfigure (\rom{1}) of \cref{fig:input_output}.
We create a rectangle for each configuration $S$ with width $y^*_S>0$ and height $u_t + B_t = v(\kappa_t) + v(\pi_t \setminus \kappa_t) \geq v(\pi_t)$. The bottom portion of the rectangle of height $u_t$ is colored dark, and the top portion of height $B_t$ is colored white. The rectangles are arranged horizontally above the interval $[0, 1]$ according to the order $\prec$.

\paragraph{Notations and Histograms for Output Distribution.}
Notice that the agent $i$ is assigned exactly one large item by Property \ref{prop:large-edge} of \cref{obs:alg-probability}. 
So our main focus is on the set $T^\rmS$ of small items assigned to $i$. 
For each $t\in(0,1]$, we define $C_t$ as the largest number $A$ such that $\Pr[v(T^\rmS)\geq A]\geq t$. Let $C_0=\lim_{t\to 0^+}C_t$.

We can create two histograms for the output distribution, one for the large items, and the other for small items. 
See the subfigure (\rom{2}) of \cref{fig:input_output}. 
Due to Property \ref{prop:marginal} of \cref{obs:alg-probability}, the histogram for large items is the same as the dark portion of the histogram for the input distribution: the height at position $t$ is $u_t$.  
So, for each set $U$ that $T^\rmS$ can take, we create a rectangle of height $v(U)$ and width $\Pr[T^\rmS = U]$, and the rectangles are sorted in descending order of $v(U)$ values from left to the right in the histogram.
In the histogram for small items, the height at position $t$ is $C_t$. 

We remark that we disregard the correlation between the large item and small items assigned to $i$, by creating two separate histograms. 
In both histograms, we sort the rectangles according to the heights. So, the large item and the small item set at position $t$ in their respective histogram may not occur simultaneously.

With the notations set up, we can relate the two quantities in \cref{lem:analysis-per-agent} with the quantities we introduced:
\begin{lemma}
$\sum_{S} y^*_S \ln v(S) \leq \int_0^1 \ln(u_t + B_t )\rmd t$.
\label{lem:input}
\end{lemma}

\begin{proof}
Consider any $t\in(0,1)$, and due to the subadditive property of the valuation function, we have $v(\pi(t))\leq u_t+B_t$.
So, $\int_{0}^{1} \ln v(\pi(t)) \rmd t \leq \int_0^1 \ln (u_t+B_t) \rmd t$.
Observe that $\sum_{S} y^*_S \ln v(S)=\int_{0}^{1} \ln v(\pi(t)) \rmd t$, as desired.
We remark that the inequality above becomes an equality when $v$ is additive. 
\end{proof}

\begin{lemma}
$\E[\ln v(T)] \geq \int_0^1 \ln(u_t + C_t)\rmd t - \ln 2$.
\label{lem:output}
\end{lemma}

\begin{proof}
Before proving the lemma, we first show the useful properties of two sequences.
\begin{claim}
Consider two non-increasing nonnegative sequences $\{a_j\}_{j=1}^{\ell}$ and $\{b_j\}_{j=1}^{\ell}$, then for any permutation $\sigma$ on $[\ell]$, we have 
\[
\sum_{j=1}^{\ell}\ln(a_j+b_{\sigma(j)})\ge\sum_{j=1}^{\ell}\ln(a_j+b_j),
\]
where $\sigma(j)$ is the index at the $j$-th position of permutation $\sigma$.
\label{claim:dual-rearrange}
\end{claim}

The \cref{claim:dual-rearrange} can be proved via the exchange argument; a proof can be found in~\cite[Theorem 2]{O54}.
Now, suppose that the \cref{claim:dual-rearrange} is correct.
We have $v(T)\ge v(k^\rmL)$ and $v(T)\ge v(T^\rmS)$ by the monotonicity of $v$, so it suffices to show that 
\begin{align}
    \E[\ln(v(k^\rmL)+v(T^\rmS))]\ge \int_0^1\ln(u_t+C_t)\rmd t. \label{inequ:sorting}  
\end{align}

To see that, we assume $\Pr[T^\rmS = \bar T, k^\rmL = \bar k]$ for any $\bar T \subseteq \smallitems$ and $\bar k \in \largeitems$ is an integer multiple of $\Delta$, for a sufficiently large integer $\Delta > 0$. We construct two multi-sets $Z^\rmL$ and $Z^\rmS$ of $\Delta$ integers as follows. For every pair $(\bar T, \bar k)$ with positive $\Pr[T^\rmS = \bar T, k^\rmL = \bar k]$, we add $\Delta\Pr[T^\rmS = \bar T, k^\rmL = \bar k]$ copies of $v(\bar T)$ to $Z^\rmL$, and $\Delta\Pr[T^\rmS = \bar T, k^\rmL = \bar k]$ copies of $v(\bar k)$ to $Z^\rmS$. Let the sequences $(a_1, a_2, \ldots, a_\Delta)$ and $(b_1, b_2, \ldots, b_\Delta)$ respectively be the sets $Z^\rmL$ and $Z^\rmS$ sorted in non-increasing order. After scaling up by $\Delta$, the left-side of \eqref{inequ:sorting} becomes $\sum_{j=1}^{\Delta}\ln(a_j+b_{\sigma(j)})$ for some permutation $\sigma$ of $[\Delta]$, and the right-side becomes $\sum_{j = 1}^{\Delta}\ln(a_j + b_j)$. \eqref{inequ:sorting} then follows from \cref{claim:dual-rearrange}.

\end{proof}
With the two lemmas, it remains for us to compare $\int_0^1 \ln(u_t + B_t)\rmd t$ with $\int_0^1 \ln(u_t + C_t) \rmd t$. 

\subsection{Comparing $\int_0^1 \ln(u_t + B_t)\rmd t$ and $\int_0^1 \ln(u_t + C_t) \rmd t$}
\label{sec:comparing}

By \ref{prop:trun:lower-bound} of \cref{lem:truncate}, we know that the truncated function ${f}$ is a lower bound of $v$. 
We shall use the lower bound ${f}$ in our analysis as the truncation gives an upper bound on the values of individual items, which allows us to apply the Chernoff-type concentration bound stated in \cref{thm:pipage-rounding}.
The following helper lemma will be useful later, which suggests a lower bound of the small items' expected value under the truncated valuation function.
\begin{lemma}
    \label{lem:expected-lower-bound}
    Let $\rho\in(0,1)$, let ${f}:=v^{(u_{\rho}+C_{\rho})}$ be the function obtained from $v$ by truncating individual values by $u_\rho + C_\rho$, as in \cref{def:truncate}. Then we have 
    \begin{align*}
    \int_{0}^{\rho} {f}(\pi(t)\setminus \kappa_t) \rmd t \geq (u_{\rho}+C_{\rho}) \cdot \int_{0}^{\rho} \frac{B_t}{u_t+C_t} \rmd t.    
    \end{align*}
\end{lemma}

\begin{proof}
We focus on any $t\in (0,\rho)$, and we have $\max_{j\in \pi(t)\setminus \kappa_t}v(j) \leq u_t \leq u_t+C_t$.
Thus, by \ref{prop:trun:lower-bound} of \cref{lem:truncate}, we have
\[
{f}(\pi(t)\setminus \kappa_t) \geq \frac{u_{\rho}+C_{\rho}}{u_t+C_t} \cdot v(\pi(t)\setminus \kappa_t)  = \frac{B_t}{u_t+C_t}\cdot (u_{\rho}+C_{\rho}).
\]
Therefore, we have
\[
\int_{0}^{\rho} {f}(\pi(t)\setminus \kappa_t) \rmd t \geq \int_{0}^{\rho} \frac{B_t}{u_t+C_t}\cdot (u_{\rho}+C_{\rho}) \rmd t=(u_{\rho}+C_{\rho}) \cdot \int_{0}^{\rho} \frac{B_t}{u_t+C_t} \rmd t. \qedhere
\]
\end{proof}

We then prove the following key lemma:
\begin{lemma}
For every $\lambda \geq 0$, we have
\[
\int_0^1 \1 \big\{\ln B_t > \ln(u_t + C_t) + \lambda\big\} \rmd t \leq (6\lambda + 1) \ce^{-\lambda}.
\]
\label{lem:ratio-key}
\end{lemma}

\begin{proof}
 We can assume that $\lambda \geq 1$ since otherwise $(6\lambda+1)\ce^{-\lambda}>1$ and the lemma holds trivially.
For each $t\in (0,1)$, we define $\theta_t=\1\big\{\ln B_t > \ln(u_t + C_t) + \lambda\big\}\in\set{0,1}$, i.e., $\theta_t=1$ if $\ln B_t > \ln(u_t + C_t) + \lambda$; otherwise, $\theta_t=0$.
It is sufficient to prove $\int_{0}^{1-\ce^{-\lambda}}\theta_t \rmd t \leq 6\lambda \ce^{-\lambda}$.
Assume towards the contradiction that $\int_{0}^{1-\ce^{-\lambda}} \theta_t \rmd t>6\lambda \ce^{-\lambda}$.
Then, we can find a $z\in (0,1-\ce^{-\lambda})$ such that $\int_{0}^{z}\theta_t \rmd t = 6\lambda \ce^{-\lambda}$.
Note that $\theta_z=1$; so, we have $\frac{B_z}{u_z+C_z} > \ce^{\lambda}$.

Now, we define the truncated function ${f}$ using the value of $u_z+C_z$, i.e., ${f}:=v^{(u_z+C_z)}$.
We aim to use the concentration bound proved in \cref{thm:pipage-rounding}, which bounds the value of small items deviating from its multilinear extension.

Let $\bx^{\umk} := (\bx^*_{ij})_{j \in \smallitems[i]}$ be fractional assignment for small items to agent $i$.
By \cref{lem:truncate}, we know that ${f}$ is a monotone submodular function.
Let $F(\cdot)$ be the multilinear extension of the function ${f}$.
Our goal is to lower bound the value of $F(\bx^{\umk})$ so that we can use the concentration bound. Let $f^+:[0, 1]^{\smallitems} \to \R_{\geq 0}$ be the concave extension of $f$, defined in \cref{def:concave}.

Now, we consider the small item part of the \eqref{Conf-LP}'s solution $(y^*_S)_{S\subseteq \smallitems}$.
By the definition of $\bx^{\umk}$ and \eqref{Conf-LP}'s constraints, $\bx^{\umk}$ and $(y^*_S)_{S\subseteq \smallitems}$ must satisfy all constraints stated in the concave extension in \cref{def:concave}, i.e., (\rom{1}) $\sum_{S\subseteq \smallitems} y^*_S \leq 1$ since \eqref{LPC:agent} of \eqref{Conf-LP}; (\rom{2}) for all small item set $S\subseteq \smallitems$, $y^*_S \geq 0$; (\rom{3}) for each small item $j\in \smallitems$, $\sum_{S:j\in S}y^*_S \leq x^{\umk}_j$.
Thus, we know that $f^+(\bx^{\umk})$ is at least as large as the expected value of small items in the LP solution, i.e.,
\[
f^+(\bx^{\umk}) \geq \int_0^1 v(\pi(t)\setminus \kappa_t) \rmd t \geq \int_0^1 {f}(\pi(t)\setminus \kappa_t) \rmd t \geq \int_0^z\theta_t \cdot {f}(\pi(t)\setminus \kappa_t) \rmd t.
\]

Applying \cref{lem:expected-lower-bound} by setting $\rho$ as $z$, we have:
\[
\int_0^z\theta_t \cdot {f}(\pi(t)\setminus \kappa_t) \rmd t \geq (u_z+C_z) \cdot \int_{0}^{z} \theta_t \cdot \frac{B_t}{u_t+C_t} \rmd t \geq (u_z+C_z)\cdot e^\lambda \cdot \int_0^z\theta_t \rmd t = (u_z + C_z) \cdot 6\lambda.
\]

By \cref{lem:concave-multilinear}, we have
\[
F(\bx^{\umk}) \geq \left(1-\frac{1}{\ce}\right)f^+(\bx^{\umk}) \geq (u_z+C_z) \cdot 6\left(1-\frac{1}{\ce}\right)\lambda.
\]
Notice that ${f}(\cdot)$ is a submodular function such that the maximum value of each single item is at most $u_z+C_z$.
Thus, the marginal value of $\frac{{f}(\cdot)}{u_z+C_z}$ is in $[0,1]$.
Thus, the concentration bound stated in \cref{thm:pipage-rounding} works for $\frac{{f}(\cdot)}{u_z+C_z}$. 
Then, we have the following inequalities:
\begin{flalign*}
&& &\quad \Pr[v(T^\rmS) < \lambda (u_z+C_z)] \\
&& &\leq \Pr[{f}(T^\rmS) < \lambda (u_z+C_z)] && [{f}(T^\rmS)\leq v(T^\rmS)] \\
&& &= \Pr\left[\frac{{f}(T^\rmS)}{u_z+C_z} < \lambda \right] \\
&& &=\Pr\left[\frac{{f}(T^\rmS)}{u_z+C_z} < \left( 1-\frac{5\ce-6}{6(\ce-1)} \right) \left(6(1-\frac{1}{\ce})\lambda\right) \right] \\
&&  &\leq \exp\left( -\frac{(\frac{5\ce-6}{6(\ce-1)})^2 \cdot 6(1-\frac{1}{\ce})\lambda}{2} \right) && \text{[\cref{thm:pipage-rounding} with $\delta=\frac{5\ce-6}{6(\ce-1)},\mu\geq 6(1-\frac{1}{\ce})\lambda$}] \\
&& &\leq \ce^{-\lambda}.
\end{flalign*}

Thus, with probability at least $1-\ce^{-\lambda}$, we have $v(T^\rmS)\geq \lambda(u_z+C_z) > u_z+C_z$ since $\lambda\geq 1$.
We use $t^*$ to represent $1-\ce^{-\lambda}$.
Recall that the definition of $C_t$ is the largest $A$ such that $\Pr[v(T^\rmS)\geq A]\geq t$ holds.
Therefore, we have $C_{t^*}>u_z+C_z \geq C_z$. Notice that $z < 1 - \ce^{-\lambda} = t^*$.
This contradicts the fact that $C_t$ is non-increasing over $t$.
\end{proof}

Now, we are ready to bound the gap between the expected value of the LP solution and the solution outputted by \cref{alg:rounding}.
\begin{lemma}
    $\displaystyle \int_0^1 \big(\ln (u_t + B_t) - \ln (u_t + C_t)\big) \rmd t \leq 3+\frac{26}{\ce^3}$.
    \label{lem:ratio}
\end{lemma}
\begin{proof}
We have the following inequality:
    \begin{flalign*}
        && &\quad \int_0^1 (\ln(u_t+B_t)-\ln(u_t+C_t))\rmd t \\
        &&  &\leq \int_0^\infty \Pr\left[\frac{u_t+B_t}{u_t+C_t}>\ce^\lambda\right]\rmd\lambda 
       \quad \leq \quad \int_0^\infty\Pr\left[\frac{B_t}{u_t+C_t}>\ce^\lambda-1\right]\rmd\lambda && [u_t\leq u_t+C_t] \\ 
       &&  &\leq 3+\int_3^\infty\Pr\left[\frac{B_t}{u_t+C_t}>\ce^\lambda-1\right]\rmd\lambda 
      \quad \leq \quad 3+\int_3^\infty\Pr\left[\frac{B_t}{u_t+C_t}>\ce^{\lambda-1/\ce^3}\right]\rmd\lambda \\ 
       &&  &= 3+\int_{3-1/\ce^3}^\infty\Pr\left[\frac{B_t}{u_t+C_t}>e^\lambda \right]\rmd\lambda
       \quad \leq \quad 3+1/\ce^3+\int_3^\infty\Pr\left[\frac{B_t}{u_t+C_t}>\ce^\lambda \right]\rmd\lambda \\
      &&   &\leq 3+1/\ce^3+\int_3^\infty(6\lambda+1)\ce^{-\lambda}\rmd\lambda && \text{[by \cref{lem:ratio-key}]} \\ 
      &&   &= 3+\frac{26}{\ce^3}. &&\qedhere
    \end{flalign*}
\end{proof}

Now, we are ready to show \cref{lem:analysis-per-agent}.
\begin{proof}[Proof of \cref{lem:analysis-per-agent}]
By \cref{lem:input}, \cref{lem:output}, and \cref{lem:ratio}, we immediately obtain the following inequalities:  
\begin{align*}
\E[\ln v(T)] - \sum_{S}y^*_S \ln v(S) 
&\geq \int_{0}^{1}\ln (u_t+C_t) \rmd t - \int_{0}^{1} \ln (u_t+B_t) \rmd t - \ln 2\\
&\geq -\left(3+\frac{26}{\ce^3} + \ln 2\right).
\end{align*}
This finishes the proof of \cref{lem:analysis-per-agent}.
\end{proof}

\section{Conclusion}

We study the problem of maximizing the weighted Nash social welfare under submodular valuations and give a $(233+\epsilon)$-approximate algorithm for the problem, which is the first constant approximation for the problem.
It improves upon the previous best-known result of $O(nw_{\max})$.
Our work leaves several interesting future directions.

It would be interesting to improve the approximation ratio to a small constant.
The current best approximation ratio for unweighted Nash social welfare with submodular valuations is $(4+\epsilon)$ approximation~\cite{GHL23}. 
There is a huge gap between the weighted and unweighted cases.

It is also interesting to consider the weighted NSW problem for more general valuations, such as XOS or subadditive valuations, where the algorithm is assumed to have access to demand oracles to these valuations. 
For the weighted case with additive valuation, an $(\ce^{1/\ce}+\epsilon)$-approximation is known to exist~\cite{FL24}, matching the best-known ratio for the unweighted case. 
This work extends the constant-factor approximation to submodular valuations. 
Extending this result further beyond the submodular case remains an intriguing open question.

\newpage
\clearpage
\bibliographystyle{plain}
\bibliography{main}

\newpage
\appendix

\section{Missed Proofs from \cref{sec:prelim}}

\subsection{Proof of \cref{thm:pipage-rounding}}
\label{sec:pipage-rounding}

In this section, we prove \cref{thm:pipage-rounding}, which shows that the randomized variables produced by \cref{alg:pipage} satisfy the Chernoff-type concentration bound. 
We shall utilize the {\em concave pessimistic estimator} technique developed in \cite{HO14}. Let $\bx^* \in [0, 1]^{\bar n}$, $v:2^{[\bar n]} \to \R_{\geq 0}$, $F: [0, 1]^{\bar n} \to \R_{\geq 0}$, $\mu = F(\bx^*)$, $\bx$, $U$ and $\delta \in (0, 1)$ be as defined in the theorem. 

We shall use the same concave pessimistic estimator as~\cite{HO14}:
\[
g_{t,\theta}(\bx):=\ce^{-\theta t}\cdot \E_{V \sim \cD(\bx)} \left[ \ce^{\theta v(V)} \right], \forall \bx \in [0, 1]^{\bar n}, 
\]
where $t:=(1-\delta)\mu, \theta:=\ln (1-\delta) < 0$ and $\cD(\bx)$ denotes the product distribution over $2^{[\bar n]}$ where $\Pr_{V \sim \cD(\bx)}[i \in V] = x_i$ for every $i \in [\bar n]$.

\begin{lemma}[\cite{HO14}]
The concave pessimistic estimator has the following properties:
\begin{enumerate}[label=(\ref{lem:est}\ablue{\alph*}),leftmargin=*,align=left]
    \item for every integral $\bx \in \{0, 1\}^{[\bar n]}$ and the set $U = \{i \in [\bar n]: x_i = 1\}$, we have ${\bf1}\{v(U) \leq t\} \leq g_{t, \theta}(\bx)$;
    \label{prop:est:bridge}
    \item $g_{t,\theta}(\bx^*) \leq \exp (-\delta^2\mu/2)$;
    \label{prop:est:bound}
    \item the function $g_{t,\theta}$ is concave on the direction $(\e_a-\e_b)$ for any $a,b\in[\bar{n}]$;
    \label{prop:est:concave}
    \item the function $g_{t,\theta}$ is linear on the direction $\e_a$ for any $a\in[\bar{n}]$.
    \label{prop:est:linear}
\end{enumerate}
\label{lem:est}
\end{lemma}

One can imagine that $g_{t,\theta}$ estimates the probability that a bad event $(v(U)\leq t)$ happens.
We remark that \ref{prop:est:bridge} of \cref{lem:est} is followed by the definition of the concave pessimistic estimator.
The \ref{prop:est:bound} of \cref{lem:est} is the standard proof for Chernoff bound.
In~\cite[Claim B.1.]{HO14}, they showed that $h(V):=e^{\theta f(V)},\forall V\subseteq [\bar{n}]$ is a supermodular function if $f:2^V \to \R$ is submodular. 
Thus, $g_{t,\theta}$ the multilinear extension of the supermodular function $e^{\theta v}$. 
It has been shown in~\cite{calinescu2011maximizing} that $F(\cdot)$ is convex on the direction $(\e_a-\e_b)$ if $F$ is a multilinear extension of a submodular function.
Thus, \ref{prop:est:concave} holds by noting that the negation of a supermodular function is submodular. 
Since $g_{t,\theta}(\cdot)$ is a multilinear extension, $g_{t,\theta}(\cdot)$ is linear on any direction $\e_a$.
This proves \ref{prop:est:linear}.

Now, we show that our rounding algorithm has a nice property (\cref{lem:concave}), which is crucial for \cref{thm:pipage-rounding}.
\cref{lem:concave} suggests that the probability of bad events (the randomized set has a low value) occurring does not increase as \cref{alg:pipage} runs.

\begin{lemma}
Given any function $f:[0,1]^{\bar{n}}\to\R_{\geq 0}$ such that (\rom{1}) $f$ is concave on the direction $(\e_a-\e_b)$ for any $a,b\in[\bar{n}]$ and (\rom{2}) $f$ is concave on the direction $(\e_c)$ for any $c\in[\bar{n}]$, then $\E[f(\bx)]\leq f(\bx^*)$. 
\label{lem:concave}
\end{lemma}

\begin{proof}[Proof of \cref{thm:pipage-rounding}]
Combining \cref{lem:est} and \cref{lem:concave}, we have the following inequalities:
\[
\Pr[v(U)\leq t] \stackrel{(\rom{1})}{\leq} \E[g_{t,\theta} (\bx)] \stackrel{(\rom{2})}{\leq} g_{t,\theta} (\bx^*) \stackrel{(\rom{3})}{\leq} \exp(-\delta^2\mu/2).
\]
The inequality (\rom{1}) is due to \ref{prop:est:bridge} of \cref{lem:est}, by taking the expectation over the distribution $\bx$ produced by \cref{alg:pipage} on both sides.
The inequality (\rom{2}) is due to \cref{lem:concave}, \ref{prop:est:concave} and \ref{prop:est:linear}. 
The inequality (\rom{3}) is due to \ref{prop:est:bound} of \cref{lem:est}.
Since $t=(1-\delta)\mu$, we have the desired concentration bound for items selected by \cref{alg:pipage}.
\end{proof}

We remark that similar to the standard Chernoff bound, \cref{thm:pipage-rounding} holds for cases where we only know a lower bound of the actual expected value.

\begin{proof}[Proof of \cref{lem:concave}]
The proof is similar to the pipage rounding proof. Fix an iteration in \cref{alg:pipage}. Let $\bx^{0}$ be the vector $\bx$ at the beginning of the iteration, and $\bx^1$ be the vector at the end. 
If \cref{alg:pipage} uses Operation 1 in this step, 
then $\bx^{1}$ is $\bx^{0}+\delta_2 \e_a$ with probability $\frac{\delta_1}{\delta_1+\delta_2}$, and $\bx^{0}-\delta_1 \e_a$ with probability $\frac{\delta_2}{\delta_1+\delta_2}$. 
Since $f$ is concave on the direction $\e_a$, we have:
\[
\E[f(\bx^{1})\mid \bx^{0}]=\frac{\delta_2}{\delta_1+\delta_2}\cdot f(\bx^{0}-\delta_1 \e_a) + \frac{\delta_1}{\delta_1+\delta_2} \cdot f(\bx^{0}+\delta_2 \e_a) \leq f\Big(\bx^{0} + \frac{-\delta_2\delta_1+\delta_1\delta_2}{\delta_1+\delta_2}\cdot\e_a\Big)=f(\bx^{0}).
\]

If \cref{alg:pipage} uses Operation 2 in this step, 
then 
$\bx^{1}$ is $\bx^{0}+\delta_2(\e_a-\e_b)$ with probability $\frac{\delta_1}{\delta_1+\delta_2}$, and $\bx^{0}-\delta_1(\e_a-\e_b)$ with probability $\frac{\delta_2}{\delta_1+\delta_2}$. 
Since $f(\cdot)$ is concave on the direction $(\e_a-\e_b)$, we have:
\[
\E[f(\bx^{1})\mid \bx^{0}]=\frac{\delta_2}{\delta_1+\delta_2} \cdot f(\bx^{0}-\delta_1(\e_a-\e_b)) + \frac{\delta_1}{\delta_1+\delta_2} \cdot f(\bx^{0}+\delta_2(\e_a-\e_b)) \leq f(\bx^*).
\]
We obtain $\E[f(\bx)] \leq f(\bx^*)$ by induction over all iterations. 
This finishes the proof of \cref{lem:concave}.
\end{proof}

\section{Missed Proofs from \cref{sec:alg}}
\subsection{Proof of \cref{lem:oracle}}
\label{sec:oracle}

The following theorem considers the \emph{submodular covering} problem:
\begin{theorem}
    \label{theorem:submodular-knapsack-dual}
    Suppose we are given an oracle to a monotone submodular function $v: 2^{[\bar n]} \to \R_{\geq 0}$, item costs $(c_i \in \Q_{>0})_{i \in [\bar n]}$, and a target value $V \geq 0$. Let $\opt = \min_{S: v(S) \geq V} \sum_{i \in S}c_i$. Then for any constant $\epsilon > 0$, there is a polynomial time algorithm that finds a set $S$ with $\sum_{i \in S} c_i \leq \opt$ and $v(S) \geq \left(1 - \frac{1}{\ce} - \epsilon\right) V$.
\end{theorem}
The theorem can be obtained using the $(1 - \frac{1}{\ce})$-approximation for the submodular maximization problem with a knapsack constraint~\cite{SV04} and binary search over $\opt$. 
We can scale $c_i$ values so that they become integers. 
The binary search technique is needed to handle the case where $c_i$'s are not polynomially bounded after scaling.  
In our algorithm, we can not afford to lose a multiplicative factor on the costs of items; therefore, we need this theorem. 

\begin{proof}[Proof of \cref{lem:oracle}]

We shall use the ellipsoid method and utilize \cref{theorem:submodular-knapsack-dual} as a subroutine of the separation oracle. 
Similar to \cite{FL24}, we can first compute the maximum weighted NSW objective $\Phi$ that can be achieved by assigning one item to each agent, using the maximum-weight bipartite matching algorithm.  
Then, we know the value of the configuration LP is between $\ln \Phi$ and $\ln(m\Phi)$. 
By making $O(\frac{\log m}{\epsilon})$ guesses, we can assume we have a number $o$ such that the value of \eqref{Conf-LP} is in $(o,o+\frac{\epsilon}{3}]$. 

We now set up a dual program \eqref{Dual-LP} of \eqref{Conf-LP}, with the objective replaced by a linear constraint. 
\begin{align}
     &&   & \tag{\text{Dual-LP}} \label{Dual-LP}\\
    &&\sum_{j\in \items} \alpha_j + \sum_{i\in \agents}\beta_i &\leq o, & \label{dual:obj} \\
    &&\sum_{j\in S}\alpha_j+\beta_i &\geq w_i\cdot\ln v_i(S), &\forall i\in \agents, S\subseteq \items \label{dual:const}\\
    &&\alpha_j &\geq 0, &\forall j\in \items 
\end{align}

As we have that the value of \eqref{Conf-LP} is in $(o, o + \epsilon]$, \eqref{Dual-LP} is infeasible. 
Given two vectors $\balpha\in\R_{\geq 0}^{\items}$ and $\bbeta\in\R^{n}$ with $\sum_{j \in M} \alpha_j + \sum_{i \in N} \beta_i \leq o$, we know \eqref{dual:const} is not satisfied for some $i \in \agents$ and $S \subseteq \items$. Our goal is to design an approximate separation oracle for the inequality above.  

By enumeration, we can guess the agent $i \in \agents$, and the item $j^* \in S$ with the maximum $v_i(j^*)$. 
We discard the items $j$ with $v_i(j) > v_i(j^*)$. As we know $j^* \in S$, we know $v_i(S) \in [v_i(j^*), mv_i(j^*)]$. 
We can then divide the interval into $O(\frac{\log m}{\epsilon})$ intervals of the form $[a, b]$ with $b \leq (1+\epsilon)a$.  
Then we guess the interval $[a, b]$ where $v_i(S)$ resides in.  
Finally, we apply \cref{theorem:submodular-knapsack-dual}, with item costs $\alpha_j$ and target value $a$. 
Then, we find a set $S' \ni j^*$ such that
\[
\sum_{j\in S'}\alpha_j \leq \sum_{j \in S}\alpha_j \qquad \text{and} \qquad v_i(S') \geq \big(1 - \frac{1}{\ce} - \epsilon\big)a \geq \big(1 - \frac{1}{\ce} - \epsilon\big)\cdot \frac{v_i(S)}{1 + \epsilon}.
\]

As \eqref{dual:const} is infeasible for $(i, S)$, we have 
\begin{align*}
    \sum_{j \in S'} \alpha_j +\beta_i \leq \sum_{j \in S}\alpha_j +\beta_i < w_i\cdot \ln v_i(S) \leq w_i \cdot \ln \Big(\big(\frac{\ce}{\ce-1} + O(\epsilon)\big) \cdot v_i(S')\Big).
\end{align*}
Using the same argument as in \cite{FL24}, we conclude that we can solve \eqref{Conf-LP} up to an additive error of $\ln \left(\frac{\ce}{\ce-1}+O(\epsilon)\right) + \epsilon$. 
We can scale $\epsilon$ by a constant at the beginning so that the additive error term is at most $\ln\left(\frac{\ce}{\ce - 1} + \epsilon\right)$. 
\end{proof}

\end{document}